\documentclass[fleqn,usenatbib]{mnras}

\usepackage{newtxtext,newtxmath}

\usepackage[T1]{fontenc}

\DeclareRobustCommand{\VAN}[3]{#2}
\let\VANthebibliography\thebibliography
\def\thebibliography{\DeclareRobustCommand{\VAN}[3]{##3}\VANthebibliography}

\usepackage{graphicx}	
\usepackage{amsmath}	

\title[Periodic nulling in PSR~B0751+32]{A study of periodic nulling in PSR~B0751+32 
with FAST}

\author[W. M. Yan et al.]{
W. M. Yan,$^{1,2}$\thanks{E-mail: yanwm@xao.ac.cn (WMY)}
N. Wang,$^{1,2}$\thanks{E-mail: na.wang@xao.ac.cn (NW)}
F. F. Kou,$^{1,2}$
Z. Y. Liu,$^{1,2}$
J. P. Yuan,$^{1,2}$
Z. G. Wen,$^{1,2}$
S. N. Sun,$^{1,2}$
\newauthor
M. Y. Zou,$^{1,3}$
Y. R. Wen$^{1,3}$
and X. J. Chen$^{1,3}$
\\
$^{1}$State Key Laboratory of Radio Astronomy and Technology, 
Xinjiang Astronomical Observatory, CAS, 150 Science 1-Street, 
Urumqi, Xinjiang, 830011, P. R. China\\
$^{2}$Xinjiang Key Laboratory of Radio Astrophysics, 150 Science 1-Street, 
Urumqi, Xinjiang 830011,  China\\
$^{3}$University of Chinese Academy of Sciences, Beijing 100049, China\\
}

\date{Accepted XXX. Received YYY; in original form ZZZ}

\pubyear{2015}

\begin{document}
\label{firstpage}
\pagerange{\pageref{firstpage}--\pageref{lastpage}}
\maketitle

\begin{abstract}
We report new results from a nulling study of PSR~B0751+32 (PSR J0754+3231), 
observed at 1250~MHz with the Five hundred meter Aperture Spherical radio Telescope (FAST). 
Our analysis confirms the presence of periodic nulling in this pulsar. Using the 
recently developed mixture model method, we obtained a nulling fraction (NF) of $35.1\% \pm 0.6\%$. 
Three independent approaches were employed to estimate the nulling periodicity, and the results 
reveal significant temporal evolution of the modulation both within individual observations and 
across different sessions. 
The pulsar exhibits an asymmetric two-component mean pulse profile, with the leading 
component brighter and narrower than the trailing one. Pulse energy analysis shows that both 
components remain stable immediately after the onset of the burst state, but subsequently undergo 
a progressive decline, with the trailing component most severely affected prior to burst 
termination. Notably, no evidence of the previously reported subpulse drifting was detected in 
our data. 
Our results challenge previous models that ascribed periodic nulling to purely geometric 
effects.
\end{abstract}




\section{Introduction} \label{sec:intro}

Pulsars have been observed to exhibit emission variability over an exceptionally wide 
range of timescales, from nanoseconds to years (e.g., \citealt{hkw+03,bmsh82,lhk+10}). 
Certain emission phenomena display well-defined 
periodicity, such as subpulse drifting, periodic non-drifting amplitude modulation, 
and periodic nulling, which can serve as powerful probes of pulsars’ emission 
geometry and magnetospheric processes (e.g., \citealt{bmm+19,bmm20a,bmm20}). 
Understanding such periodic behaviour provides valuable constraints on the physical 
conditions and plasma dynamics within the magnetosphere.

Subpulse drifting is a quasi-periodic modulation where subpulses shift in phase each 
rotation, described by $P_{2}$ and $P_{3}$ \citep{bac70b,es02}, and generally attributed 
to rotating sub-beam patterns above the polar cap. It occurs mainly in conal emission and 
can switch between discrete drift modes (e.g., \citealt{mbt+17,rwyw22}), often linked to 
other variability such as nulling or mode changing \citep{vsrr03,jv04,gyy+17}, implying a 
common magnetospheric origin.

In addition to drifting, some pulsars show periodic intensity variations without phase 
motion (e.g., \citealt{mr17}), known as periodic amplitude modulation. This behaviour, 
seen in both core and conal regions \citep{bmm+16,bmm20,ymw+19,ymw+20,kyp+21,zywy23}, 
reflects magnetospheric processes that modulate emission strength and is not restricted by 
the geometric conditions required for drifting.

Pulsar nulling is the temporary cessation of detectable radio emission (e.g., \citealt{bac70}). 
Although long considered random, some pulsars exhibit distinct periodic or quasi‑periodic 
patterns, now referred to as periodic 
nulling \citep{rw07,rw08,hr07,hr09,rwb13,bmm17,blk20}. 
Periodic nulling was initially attributed to an empty line of sight passing through gaps 
between emitting sub‑beams \citep{rw08}. Later studies showed that it occurs in both core and 
conal components, and is present even in high spin‑down–energy pulsars, suggesting a different 
underlying mechanism \citep{bmm17,bmm20}.
Periodic nulling can also be interpreted geometrically if some radiating sub‑beams become 
inactive. When the line of sight crosses these extinguished sub‑beams, nulls appear, and in a 
carousel of evenly spaced sub‑beams such inactive regions generate recurring nulls at a 
characteristic period, as first suggested by \citet{rit76}. Periodic nulling offers insight into 
global magnetospheric state changes: unlike drifting subpulses or amplitude modulation, which 
affect only parts of the profile, it represents a complete suppression of detectable emission, 
implying a large‑scale change in particle acceleration or coherence conditions.

PSR~B0751+32 was discovered during a search of the northern sky \citep{dth78}. This pulsar 
has a  characteristic age of 2.12 $\times$ $ 10^7$ yr and a spin period of 1.44 s. Previous 
studies reported that PSR~B0751+32 displays both subpulse drifting 
\citep{bac81,wes06,wse07} and periodic nulling \citep{hr09,bmm17}.
In this paper, we carry out a detailed investigation of periodic nulling of PSR~B0751+32 
with observations lasting for almost 3.5 h. The observations and data reduction procedures are 
described briefly in Section~\ref{sec:obs}, the detailed results are presented in Section~\ref{sec:results}, 
the implications are discussed in Section~\ref{sec:disc}, and our findings are summarised in Section~\ref{sec:summary}.

\section{Observations} \label{sec:obs}

\begin{table*}
	\begin{center}
	\caption{Observational parameters of PSR~B0751+32. Note that the symbols 
	$\tau_{\mathrm{samp}}$ and T$_{\mathrm{obs}}$ represent the sampling 
	interval and the duration of the observation, respectively. }\label{tab:obs}
	\begin{tabular}{ccccccc}
	\hline
	Date & Frequency & Bandwidth & No. of & $\tau_{\mathrm{samp}}$ & T$_{\mathrm{obs}}$ & No. of\\
	(yyyy-mm-dd)   & (MHz)     & (MHz)     & Channels & ($\mu$s)  & (min) & Pulses\\
	\hline
	2019-07-16   & 1250 & 500 & 4096 & 49.15 &120  &4992  \\
	2019-07-18  & 1250 & 500 & 4096 & 49.15 & 88 &3661 \\
	\hline
	\end{tabular}
	\end{center}
\end{table*}

FAST, completed in September 2016 with a maximum effective aperture of 300 m, was equipped 
in May 2018 with a 19-beam L-band receiver covering the frequency range 1050 --- 1450 MHz 
\citep{jth+20}. After a series of commissioning observations, the telescope began full scientific
operations in January 2020. During the commissioning phase, we observed PSR~B0751+32 on 2019 
July 16 and 18 using the central beam of the 19-beam receiver at a frequency of 1250~MHz. 
A polarization calibration noise signal was injected at an off-source position and recorded 
prior to each pulsar observation in order to facilitate polarization calibration.
A total of 8653 single pulses were 
collected over approximately 3.5 hours of observations. Data were acquired in PSRFITS search 
mode \citep{hvm04}, with a sampling interval of 49.15 $\mu$s and 4096 frequency channels. 
A summary of the observational details is given in Table~\ref{tab:obs}.

Using the \texttt{DSPSR} package \citep{vb11} and updated pulsar ephemerides
\footnote{\url{http://www.atnf.csiro.au/research/pulsar/psrcat/}} 
\citep{mhth05}, we de-dispersed the raw data and formed singlepulse integrations. 
Radio-frequency interference was removed with standard PSRCHIVE routines \citep{hvm04} before 
further processing. Fluctuation spectra were obtained with \texttt{PSRSALSA} \citep{wel16}, 
and polarization calibration was conducted as described by \citet{ymv+11}. The resulting data 
products provided calibrated 
Stokes parameters for subsequent analysis.

\section{Results} \label{sec:results}

The results of polarization, nulling, and subpulse drifting analyses are presented 
in this section.

\subsection{Polarization} \label{sec:poln}

\begin{figure}
	\includegraphics[width=\columnwidth]{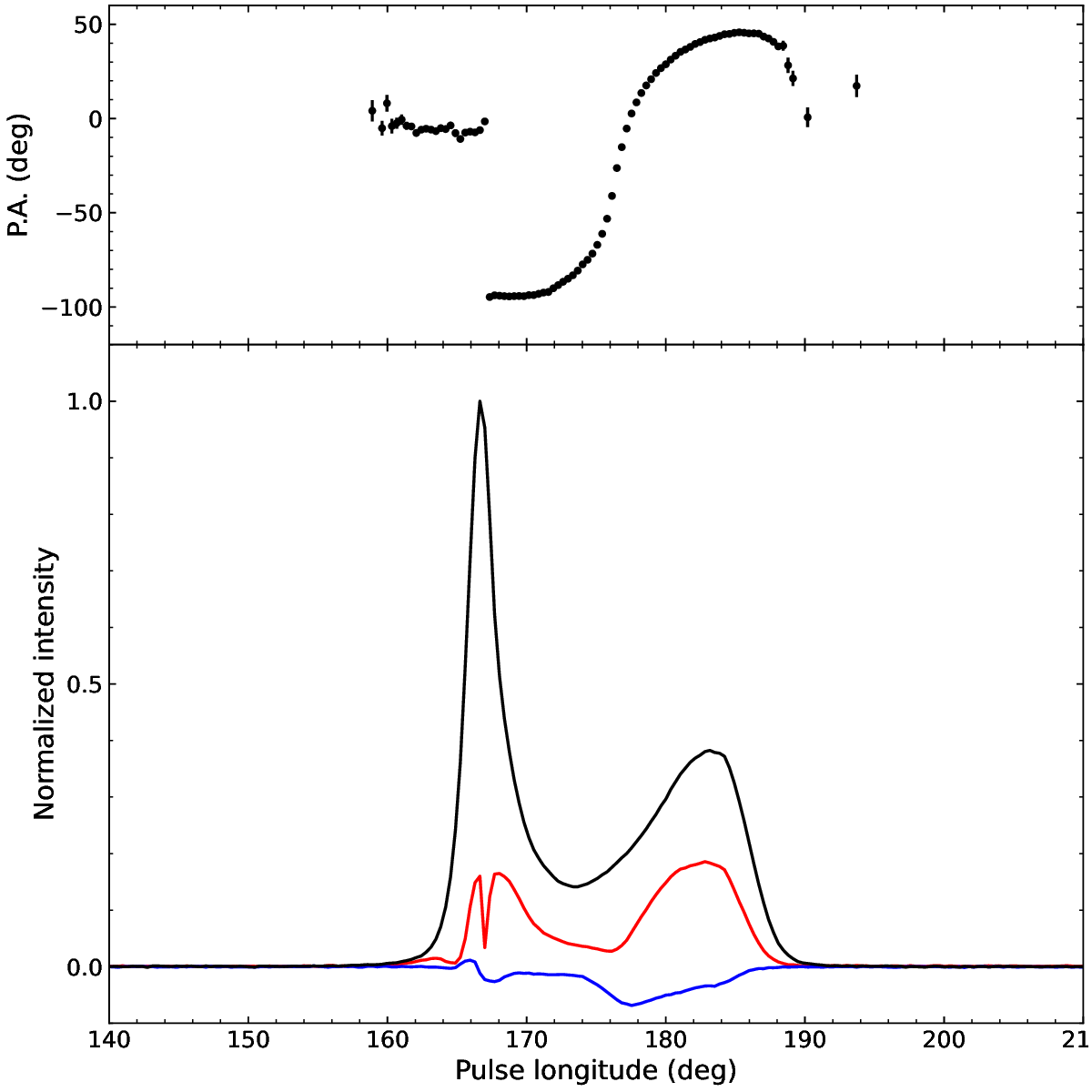}
	\caption{The polarization profile of PSR~B0751+32 derived from the 2019 July 16 
	observation. The lower panel shows the total intensity (black), linearly polarized 
	intensity (red), and circularly polarized intensity (blue). In the upper panel, 
	black dots with error bars mark the position angles of the linearly polarized 
	emission.
	}
	\label{fig:poln}
\end{figure}

Using the \texttt{RMFIT} program, we determined the RM of PSR~B0751+32 based on 
observations from 16 and 18 July 2019. The resulting weighted 
average, 5.29 $\pm$ 0.02 rad m$^{-2}$, is in close agreement with the value 
obtained by \citet{rvw+23}. Mean pulse profiles and polarization parameters for 
this pulsar are given in Figure~\ref{fig:poln}. Our results confirm and extend 
those reported in earlier studies \citep{omr19,wcl+99}. We detect a weak linear 
polarization component at the 
leading edge of the leading component of the pulse profile at a pulse longitude of 163\degr.
A clear orthogonal mode transition is observed in both linear and circular polarization 
near the peak of the pulse profile, and an orthogonal position angle (PA) transition 
occurs at a pulse longitude of 167\degr. The fractional linear polarization, 
fractional circular polarization and absolute circular polarization fraction for the 
mean profile of PSR~B0751+32 are 19.96\%, 0.17\%, and 7.26\%, respectively. 

As shown in the upper panel of Figure~\ref{fig:poln}, the PA swing in PSR~B0751+32 
resembles an S-like morphology. The rotating vector model \citep[RVM,][]{rc69a} 
fitting yields small values of $\alpha$ (magnetic inclination angle) 
and $\beta$ (impact angle), but these results should be treated 
with caution because the fit fails the significance test. 
\citet{omr19} classified PSR B0751+32 as a conal double (D) profile and derived robust geometric 
parameters: $\alpha \approx 26^\circ$, 
$\beta \approx +1\fdg0$, and a steep PA 
sweep rate $R = |{\rm d}\chi/{\rm d}\varphi| \approx 25^\circ/^\circ$, where $\chi$ is the PA 
and $\varphi$ is the pulse longitude. 
The relation $R = \sin\alpha / \sin\beta$ confirms a large $\alpha$ and a small $\beta$, fully consistent 
with our observed steep PA swing. 
Following the core/double-cone framework \citep{ran93b,ran93,omr19}, 
the core radius is $\rho_{\rm core} \approx 2\fdg45/\sqrt{P}/(2\sin\alpha) \approx 2\fdg4$, 
and the outer conal beam radius is 
$\rho_2 \approx 5\fdg75/\sqrt{P} \approx 4\fdg8$. 
These values imply that PSR B0751+32 exhibits a typical outer conal double profile, with emission 
dominated by outer conal radiation rather than core emission.

\subsection{Nulling} \label{subsec:nulling}

\begin{figure}
	\includegraphics[width=\columnwidth]{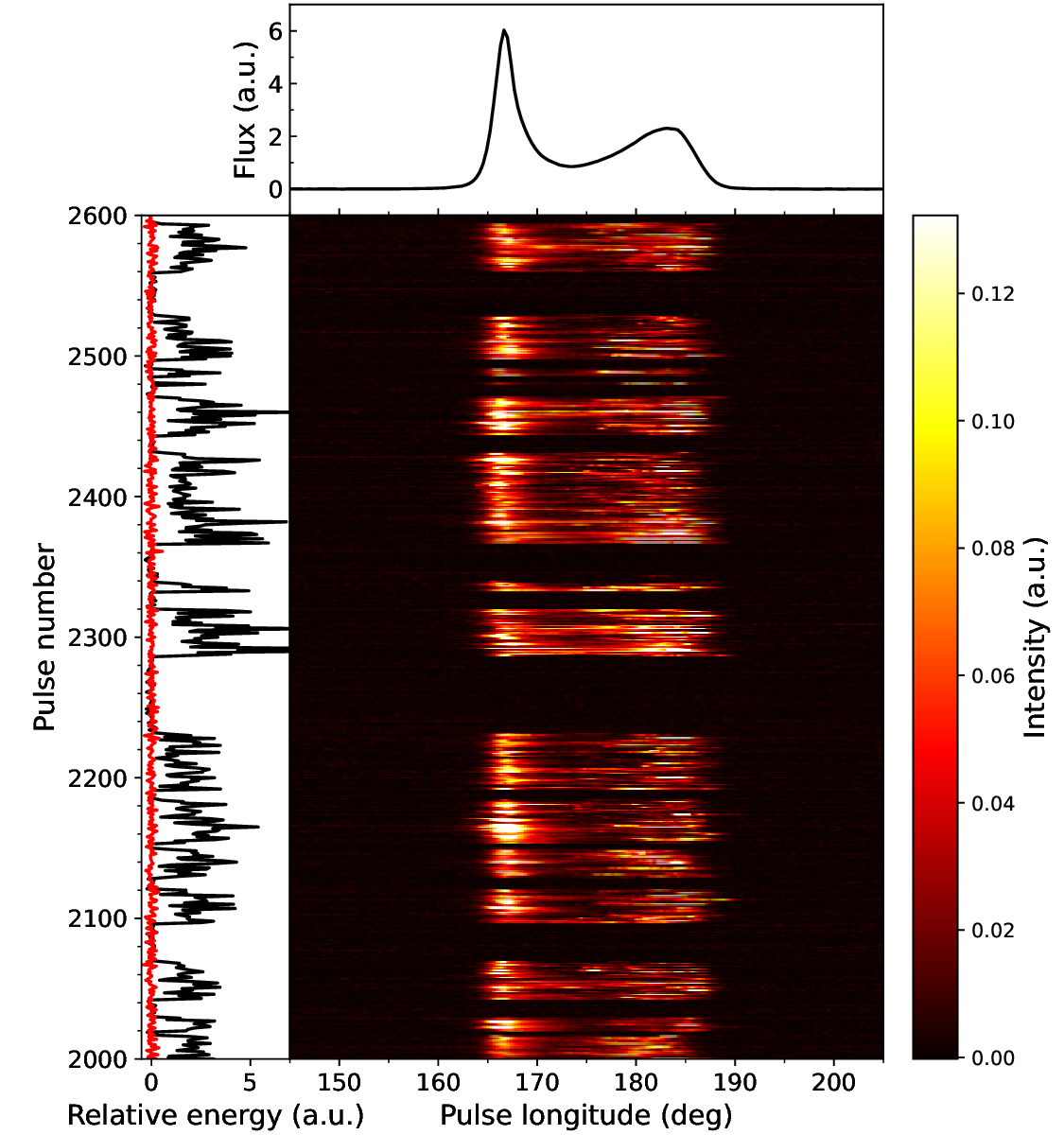}
	\caption{A single-pulse stack of 600 successive pulses (lower) 
	    from the 2019 July 16 observation and the 
		mean pulse profile (upper) of PSR~B0751+32. The left panel shows 
		the energy sequences of the on-pulse (blue) and off-pulse (red) regions.}
	\label{fig:stack}
\end{figure}

Figure~\ref{fig:stack} presents a single-pulse stack of 600 successive pulses obtained 
on 2019 July 16, in which pulse nulling is clearly evident. Motivated by this, 
we performed a detailed nulling analysis for PSR~B0751+32 in this subsection.

\subsubsection{Nulling fraction} \label{subsec:nf}

\begin{figure}
	\includegraphics[width=\columnwidth]{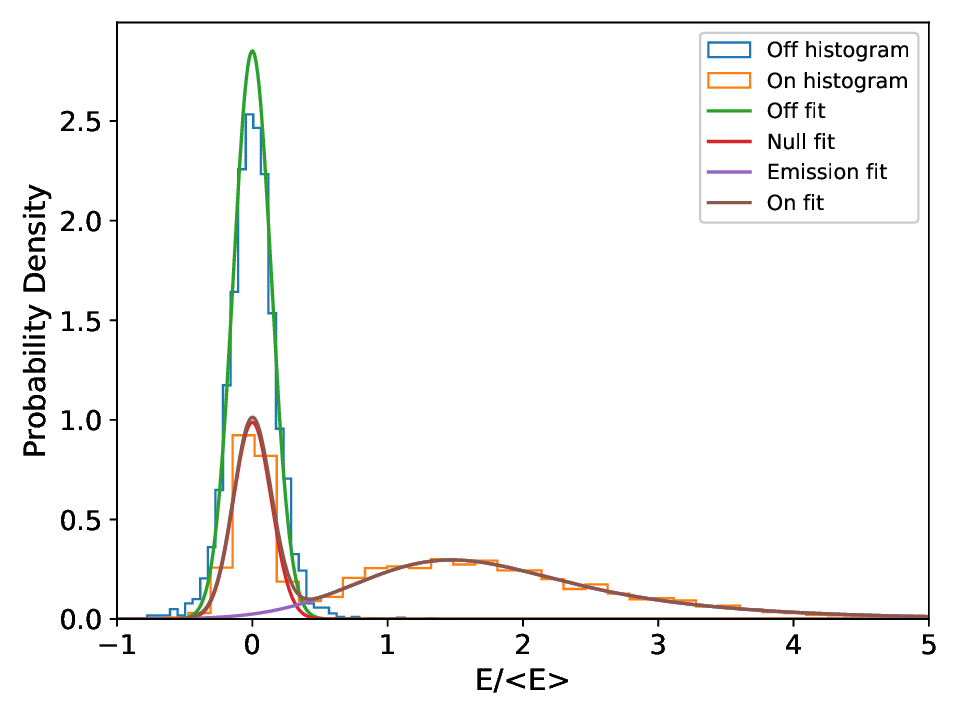}
	\caption{Pulse energy distributions for the on-pulse (orange histogram) and 
	off-pulse (blue histogram) regions from the 2019 July 16 observation. 
	The off-pulse histogram is fitted with a simple Gaussian function (green curve), 
	whereas the on-pulse histogram is fitted with a two-component mixture model, in which 
	the emission component is modeled as an exponentially modified Gaussian function 
	(purple curve) and the null component is modeled as a simple Gaussian function 
	(red curve). The brown curve represents the total fitted on-pulse model as the 
	sum of null and emission contributions.}
	\label{fig:energy_dist}
\end{figure}

\begin{table}
	\begin{center}
	\begin{minipage}{\textwidth}
				\caption{Nulling properties of PSR~B0751+32.}
					\label{tab:para}
	\end{minipage}
	\begin{tabular}{cccc}
	\hline
	&   &Burst  &Null  \\
	Date   & NF &time-scale &time-scale \\
	(yyyy-mm-dd)     & (\%)     & (period) & (period)\\
	\hline
	2019-07-16    & 34.6$\pm$0.8 &16.4$\pm$0.5  &8.2$\pm$0.1  \\
	2019-07-18  & 35.8$\pm$0.9 &14.2$\pm$0.6   &8.0$\pm$0.1 \\
    \hline
	\end{tabular}
	\end{center}
\end{table}

The on-pulse energy was determined by 
integrating the intensities within the on-pulse window, defined as the longitude 
interval in the integrated profile over which the signal-to-noise ratio exceeds 
$3\sigma$ relative to the baseline root-mean-square (RMS) noise. 
The off‑pulse energy was measured using an equally sized window away 
from the emission region. Figure~\ref{fig:energy_dist} shows the on‑pulse and 
off‑pulse energy distributions, each normalized by the mean pulse energy. The 
off‑pulse histogram is well described by a zero‑centered Gaussian, while the 
on‑pulse histogram is bimodal, with peaks at zero and at the mean burst energy, 
corresponding to the null and burst states.

The \citet{rit76} method is a commonly used approach to estimate a pulsar's NF. 
However, it may yield overestimated NF values, especially in cases where the pulsar 
has a low signal-to-noise ratio \citep{ksf+18,ask+23}. 
Here, we employ the mixture model method proposed by \citet{ksf+18} and subsequently 
expanded by \citet{ask+23}, which is publicly available via the 
\texttt{PULSAR\_NULLING} software 
package\footnote{\url{https://github.com/AkashA98/pulsar_nulling}}, to model the 
distributions of single-pulse energies and thereby estimate the NF for 
PSR~B0751+32.

As shown in Figure~\ref{fig:energy_dist}, the off-pulse histogram is fitted with a 
single Gaussian function, represented by the green curve, while the on-pulse histogram 
is described by a composite fit from a two-component mixture model, represented by 
the brown curve. The emission component is fitted with an exponentially modified 
Gaussian function, shown as the purple curve, and the null component is modeled 
with a single Gaussian function, shown as the red curve. 
The corresponding point estimates 
of the NF are provided in Table~\ref{tab:para}. 
The derived NF values for the two observations are in good agreement within the measurement 
uncertainties. The average NF is found to be $35.1\% \pm 0.6\%$.

\subsubsection{Burst and null timescales} \label{subsec:timescale}

\begin{figure}
	\includegraphics[width=\columnwidth]{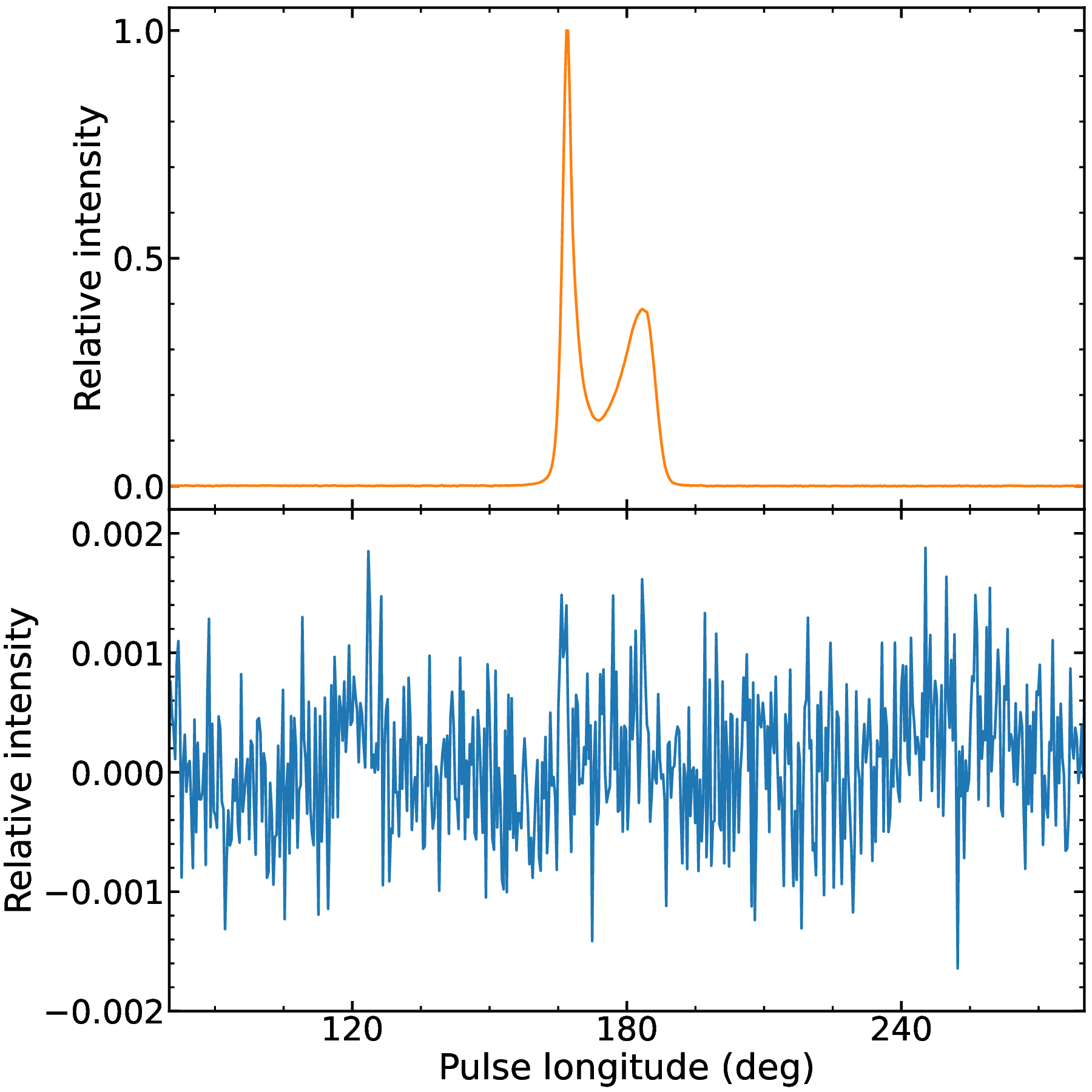}
	\caption{Mean pulse profiles of null (lower) and burst (upper) pulses 
	for PSR~B0751+32.}
	\label{fig:state_prf}
\end{figure}

\begin{figure}
	\includegraphics[width=\columnwidth]{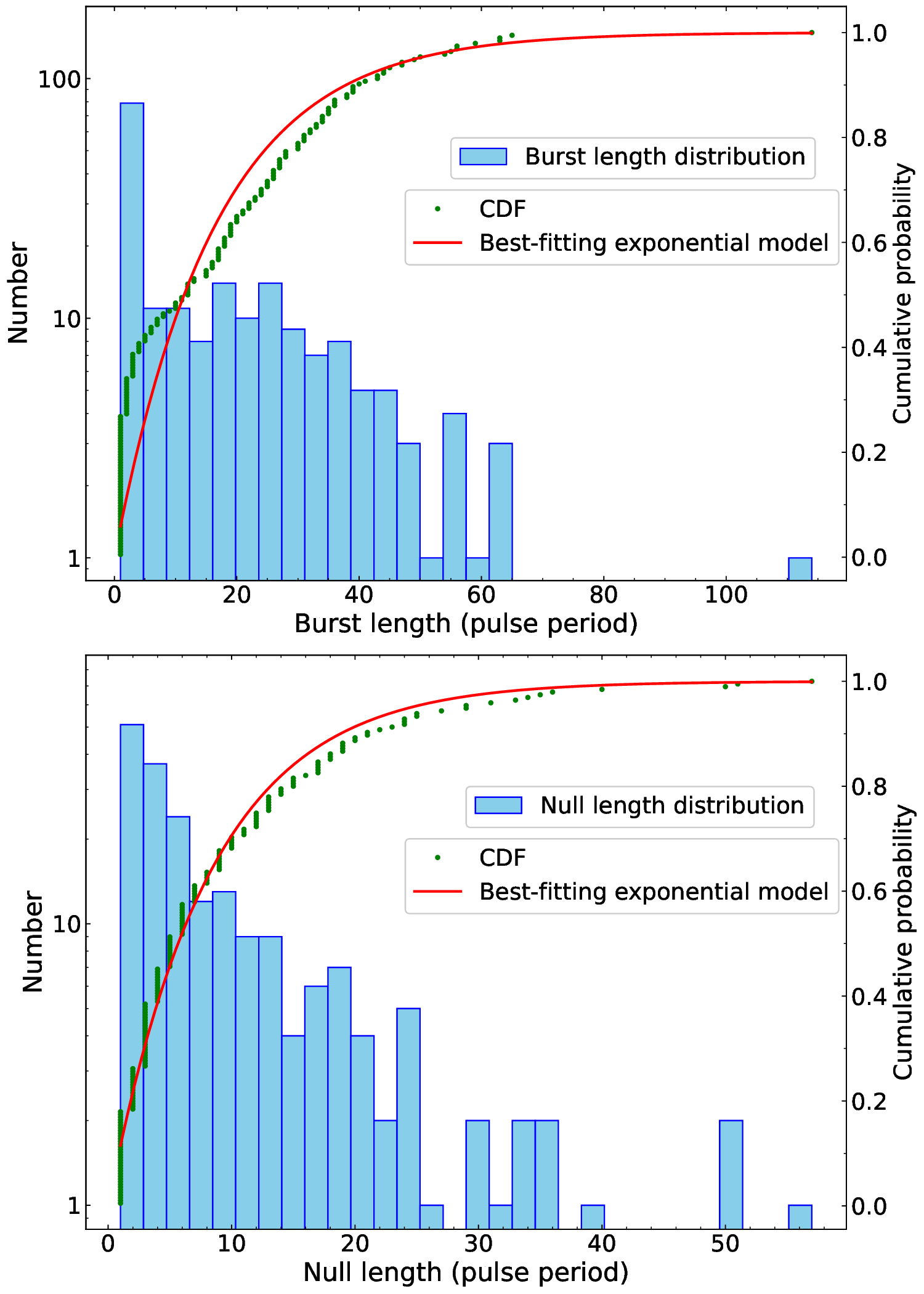}
	\caption{Distributions of burst (upper) and null (lower) lengths for PSR~B0751+32. 
	The green dots represent the corresponding CDF. 
	The red curves show the best-fitting exponential model to the CDF.
}
	\label{fig:length}
\end{figure}

Using the \texttt{PULSAR\_NULLING} package, we calculated the nulling probability 
for each individual pulse. 
A threshold of 0.5 is commonly adopted to divide 
all individual pulses into two categories \citep{ask+23}: pulses with nulling probability 
$>$~0.5 are classified as nulls, while those below the threshold are identified as bursts. 
Based on this threshold, all individual pulses of PSR~B0751+32 
were classified into two categories, and their mean pulse profiles are shown in Figure~\ref{fig:state_prf}. 
The lower panel reveals that the mean profile of nulls exhibits no detectable emission.

From the burst-null sequence derived, we constructed the distributions of null 
and burst lengths (Figure~\ref{fig:length}). \citet{gjk12} proposed that null and burst 
lengths follow an exponential distribution, characteristic of a Poisson process. The cumulative distribution function (CDF), $F(x)$, 
for such a process is given by
\begin{equation}\label{eq:exp_cdf}
    F(x) = 1 - e^{-x/\tau},
\end{equation}
where $\tau$ denotes the characteristic timescale of the stochastic process.
A least-squares fit to this model provides estimates of the characteristic null and burst 
timescales. Figure~\ref{fig:length} illustrates the best-fitting exponential models to the CDFs 
of null and burst lengths, obtained for the 2019~July~16 observation.  
The characteristic null and burst timescales derived from the two observations are 
presented in Table~\ref{tab:para}. The observed variation is likely due to differences in 
the pulse sequence lengths between the two datasets. 

\subsubsection{Nulling periodicity} \label{subsec:periodicity}

\begin{figure}
	\includegraphics[width=\columnwidth]{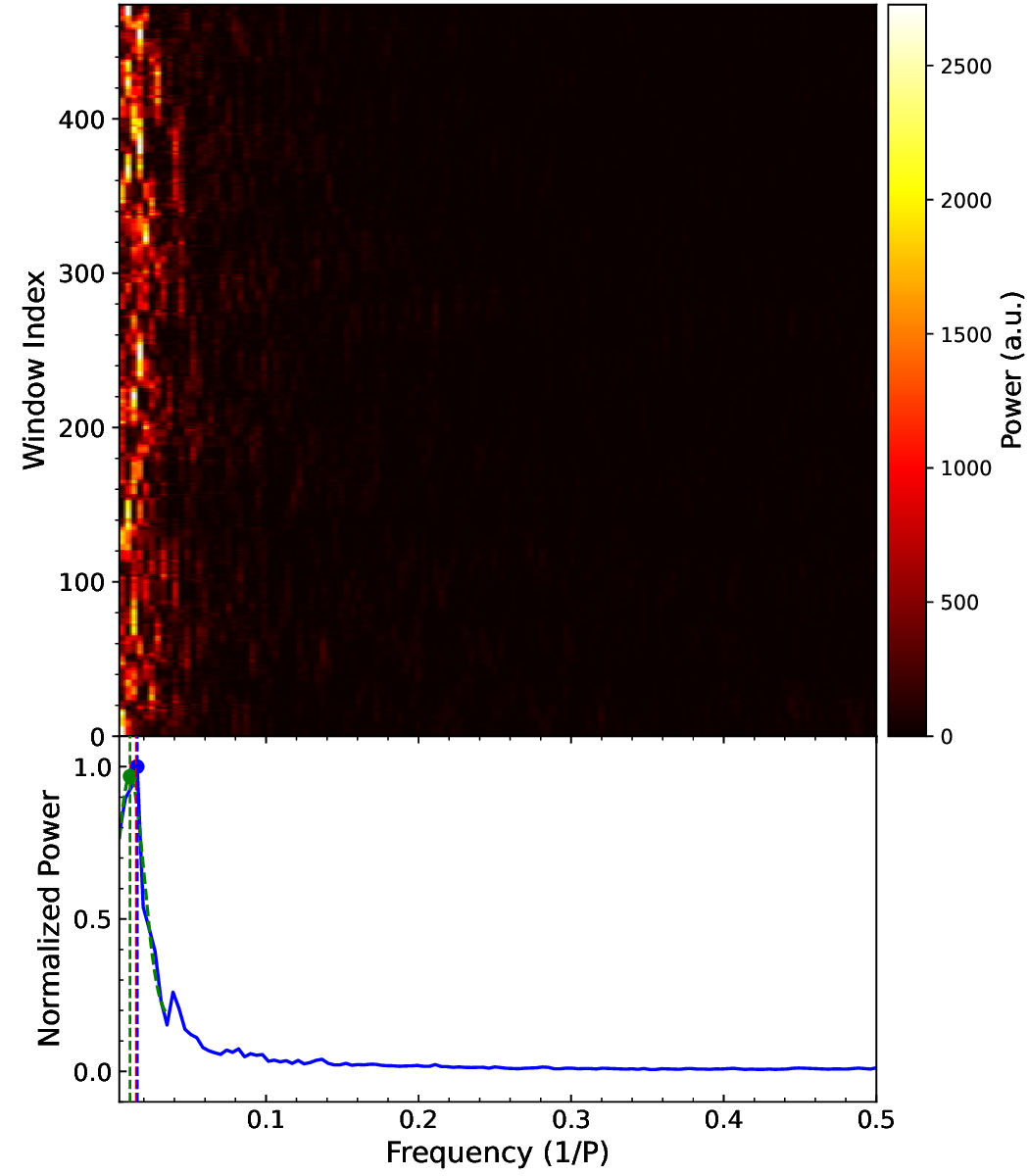}
	\caption{Sliding-window DFT analysis of the 0-1 sequence. The top panel presents the 
	dynamic power spectrum versus the DFT window index, while the bottom panel gives 
	the time-averaged normalized power spectrum. The Gaussian fit to the modulation feature 
	is shown as a green dashed curve. The frequency peak is determined using three methods: 
	bin-interval (blue dot), centroid (red dot, overlapped by the blue dot), and Gaussian 
	fitting (green dot).
}
	\label{fig:state_spec}
\end{figure}

\begin{table}
	\begin{center}
	\begin{minipage}{\columnwidth}
				\caption{Nulling periodicity of PSR~B0751+32 estimated using 
				different methods.}
					\label{tab:period}
	\end{minipage}
	\begin{tabular}{cccc}
	\hline
	Date   & Bin-interval & Centroid &Gaussian ﬁtting \\
	(yyyy-mm-dd)   & (period)     & (period)     & (period) \\
	\hline
	2019-07-16   & 64$\pm$16 & 66$\pm$23 &92$\pm$9  \\
	2019-07-18  & 85$\pm$28 & 79$\pm$25 &85$\pm$15   \\
    \hline
	\end{tabular}
	\end{center}
\end{table}

Following the method of \citet{gyy+17}, we analyzed our observations from 16 and 
18~July~2019 to test for nulling periodicity. We labeled burst pulses as 1 and null 
pulses as 0, then applied a discrete Fourier transform (DFT) to the resulting binary 
sequence using a 256-pulse sliding window, with the window shifted by 10 pulses at each step. 
The results from the 16~July~2019 observation are presented 
in Figure~\ref{fig:state_spec}.
The dynamic power spectrum in the upper panel of Figure~\ref{fig:state_spec} displays diffusely 
distributed yellow and red stripes, which reveal the temporal variation characteristics of 
the nulling periodicity. The lower panel of Figure~\ref{fig:state_spec} shows the time-averaged 
normalized power spectrum. We used three methods to determine the frequency peak: bin-interval, 
centroid, and Gaussian fitting. 

In the bin‑interval method, the peak in the time‑averaged DFT spectrum is taken as the 
characteristic frequency, and the period is obtained from its inverse. The period uncertainty 
follows directly from the spectral frequency resolution. 
We also applied the centroid method \citep{bmm+16} to identify the peak frequency in the 
time‑averaged, normalized power spectrum. The spectrum was divided into five segments to determine 
the baseline from the one with the lowest RMS. Any structure with at least three consecutive points 
exceeding the baseline by more than five RMS was treated as a candidate peak. The peak frequency was 
then obtained from the centroid of this region, and its width (FWHM) was used to estimate the RMS 
and the uncertainty in the peak frequency. The resulting peak frequency and its error were finally 
converted to the period and period uncertainty following the same procedure as in the bin‑interval method. 
To refine the frequency and period estimates beyond the DFT bin resolution, we fitted a 
Gaussian to the spectral peak. A narrow window around the maximum in the DFT spectrum was extracted and 
fitted with a Gaussian using non‑linear least squares. The fitted center frequency and its uncertainty were 
then converted to the period and its error through standard error propagation. 

The periods derived from the three independent methods for the two observing sessions are 
summarized in Table~\ref{tab:period}. 
For the 2019 July 16 observation, the bin‑interval and centroid methods give consistent, short periods, 
while the Gaussian fit yields a slightly larger value, likely due to sensitivity to peak shape. For the 2019 
July 18 observation, all three methods agree within uncertainties, indicating a stable and well‑defined peak. 
The strong cross‑method consistency, especially in the second epoch, supports the reliability of the measured 
periods. The clear difference between the two epochs further indicates temporal variability in the 
nulling periodicity. 

\subsubsection{Variability of pulse energy during emission state transitions} 
\label{sec:variability}

\begin{figure}
	\includegraphics[width=\columnwidth]{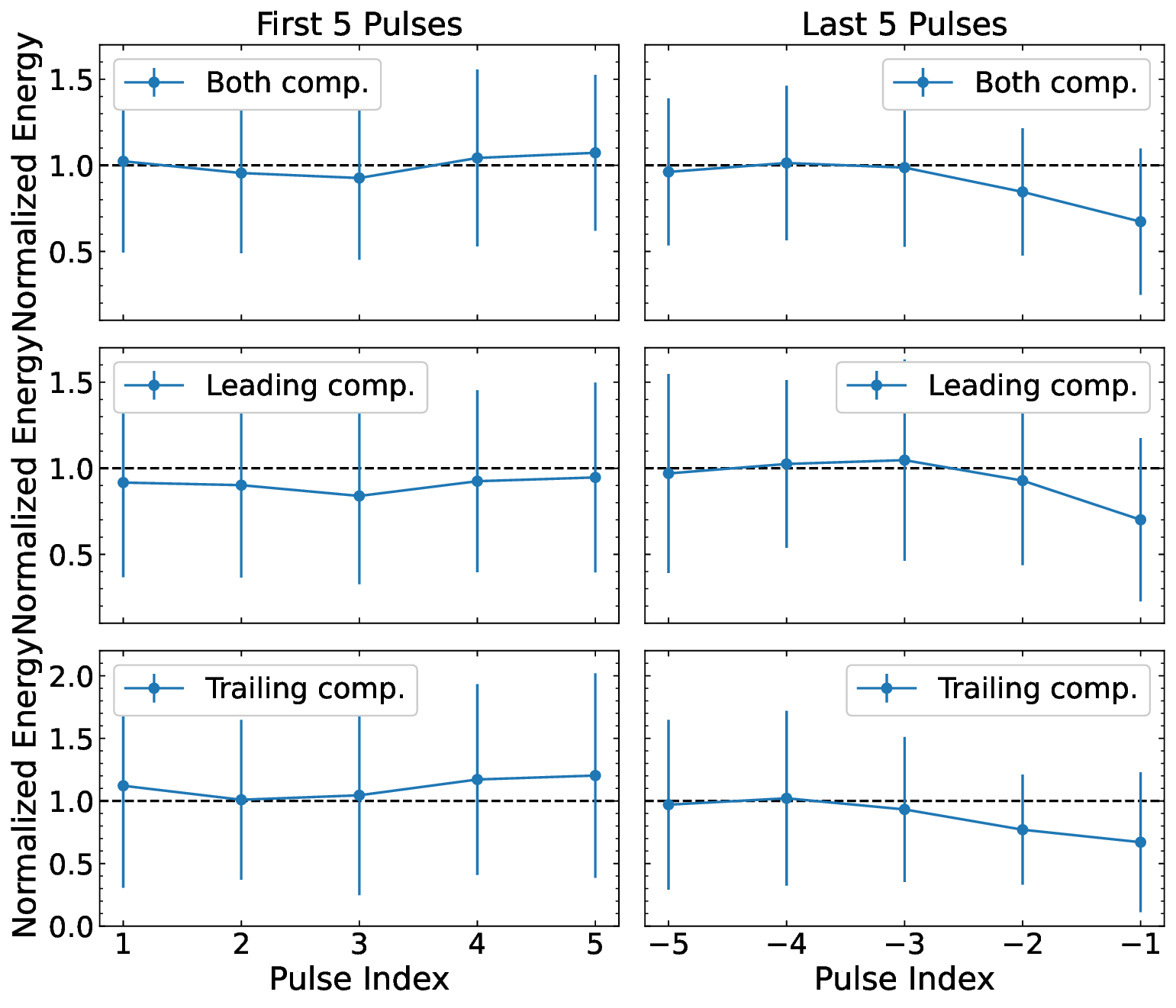}
	\caption{Relative mean pulse energy variation for the first five pulses (left column) 
	and the last five pulses (right column) during the burst state. The top, middle, 
	and bottom rows show the pulse energy variation for the whole on-pulse window, the 
	leading component, and the trailing component, respectively. 
	The horizontal dashed lines denote the mean energy of the pulse components in the burst state.}  
	\label{fig:var_around_null}
\end{figure}

\begin{figure}
	\includegraphics[width=\columnwidth]{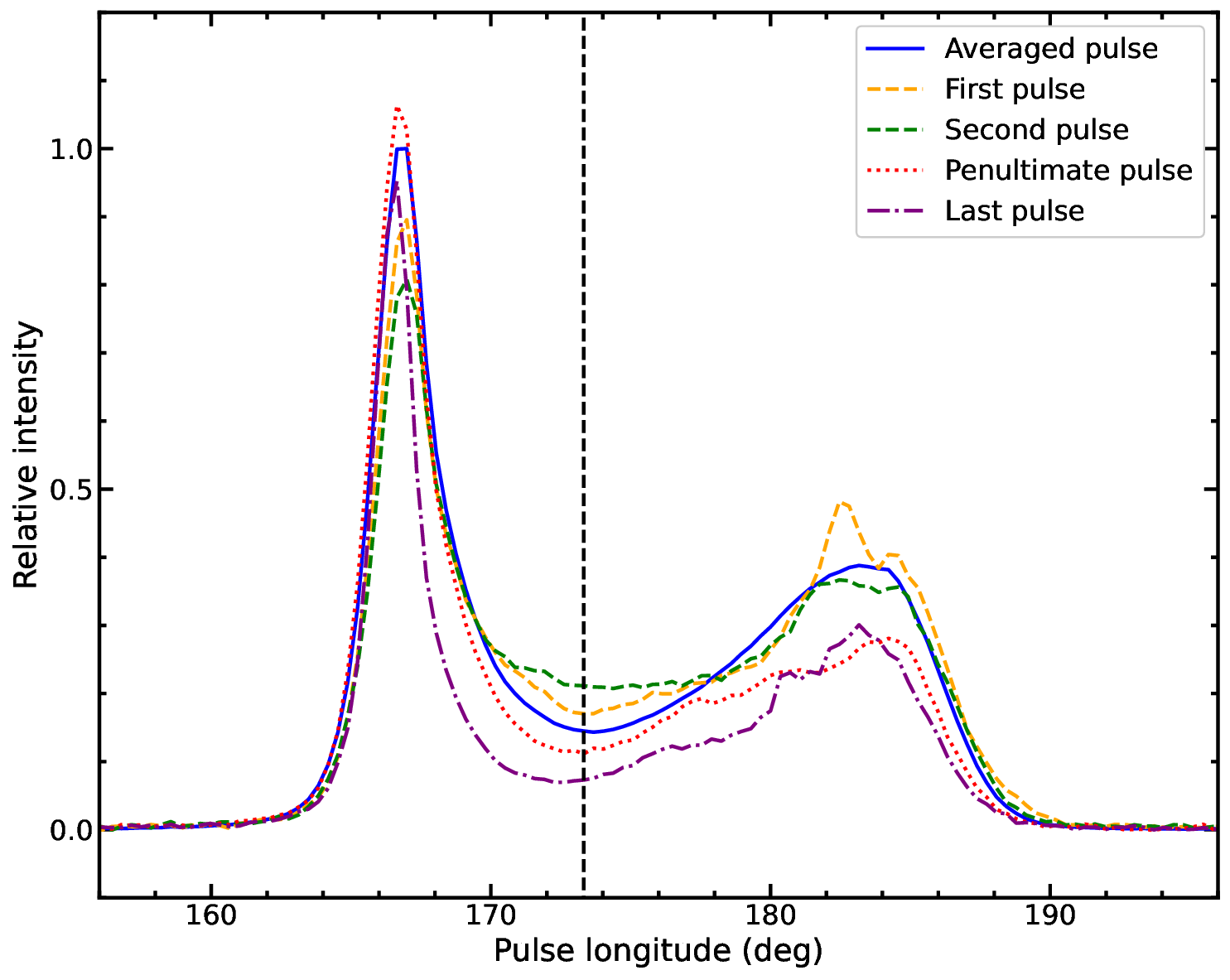}
	\caption{Mean profiles of the first (blue), second (green), penultimate (red), and 
	last (purple) pulses during the burst state. For comparison, the mean profile obtained 
	from all burst pulses (orange) is also presented. The vertical black dashed line marks 
	the boundary between the two profile components.}
	\label{fig:em_gt5_profile}
\end{figure}

In this subsection, pulse energies are analysed to characterize their temporal evolution over the 
onset and termination of the burst state. For burst states lasting at least ten pulses, pulse 
energy sequences are normalized to their respective mean energies. The average energies 
of the first five and the last five pulses were obtained by averaging, at each corresponding pulse 
position, across all such burst states. 

Figure~\ref{fig:var_around_null} shows the relative mean pulse energy variations measured for the first 
five pulses (left panels) and the last five pulses (right panels) within the burst state. 
In the first five pulses, the normalized energies of both the leading and trailing components, as well as 
the total emission, remain close to the burst-state mean. Variations are modest and within measurement 
uncertainties, indicating relatively stable radiative output at the onset of the burst state. In 
contrast, the last five pulses display a gradual decline in normalized energy, most notably in the trailing 
component. The leading 
component also exhibits a downward trend, albeit with smaller magnitude, suggesting that both components 
contribute to the fading emission, though the trailing component is primarily responsible for the observed 
reduction in total pulse energy. This decreasing trend toward the termination of the burst state may 
reflect changes in the underlying emission physics.

Figure~\ref{fig:em_gt5_profile} compares the mean profiles of selected pulses at different stages of 
the burst state lasting for at least five periods. The first pulse (orange dashed line) and the second 
pulse (green dashed line) closely match the overall burst‑state average profile (blue solid line). 
In contrast, the penultimate pulse (red dotted line) and the last pulse (purple dot‑dashed line) show 
significantly reduced intensity in the trailing component. This decreasing trend agrees well with that shown in the right panels of Figure~\ref{fig:var_around_null}.

\subsection{Subpulse drifting} \label{sec:drifting}

\begin{figure*}
	\includegraphics[width=\textwidth]{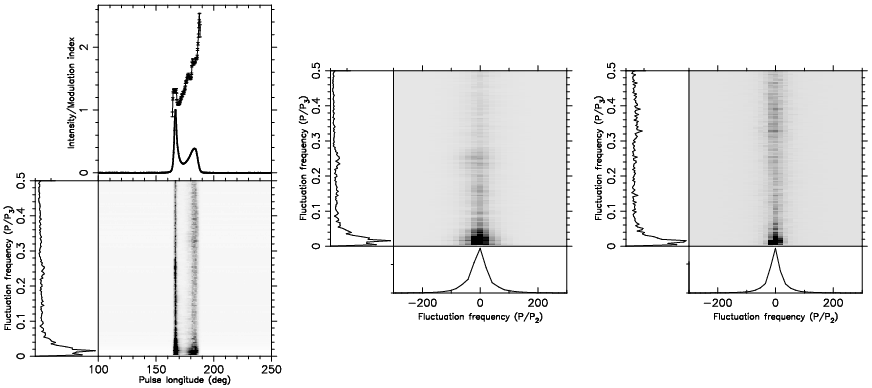}
	\caption{Results of fluctuation analysis for the 2019 July 16 observation. The upper 
	panel in the left column shows the mean pulse profile (solid line) together with the 
	longitude-resolved modulation index (points with error bars). Below, the LRFS is 
	displayed along with a side panel showing the horizontally integrated power. For 
	the leading component (middle column) and the trailing component (right column), 
	the 2DFS is plotted, with side panels displaying the horizontally (left) and 
	vertically (bottom) integrated power.}
	\label{fig:2dfs}
\end{figure*}

The ultra‑high sensitivity 
of FAST offers an opportunity to detect the subpulse drifting pattern in PSR~B0751+32, 
should it exist.
To investigate the subpulse modulation properties of PSR~B0751+32, we computed the 
longitude-resolved fluctuation spectrum \citep[LRFS,][]{bac70b} and the two-dimensional 
fluctuation spectrum \citep[2DFS,][]{es02} for the two observations using \texttt{PSRSALSA}. 
For detailed descriptions of these analysis techniques, the reader is referred to \citet{wes06}.

Analysis of the LRFS (Figure~\ref{fig:2dfs}) reveals that, for both the leading and trailing 
components, the spectral peak occurs at 0.0156~cycles\,per\,period\,(cpp), 
indicating that both components have the same $P_3 = 63.9 \pm 0.3\ P$, 
where $P$ denotes the pulsar rotation period. 
This value is consistent with the nulling periodicity listed in Table~\ref{tab:period}.
The 2DFS exhibits perfect symmetry about the vertical axis for both the leading and trailing 
components, indicating that subpulses in successive pulses do not show an average drift toward 
later or earlier pulse longitudes. Even after excluding all null-state pulses and analysing 
only burst pulses, the 2DFS remains fully symmetric, demonstrating that periodic nulling does not 
mask any drifting signature. Consequently, we conclude that no subpulse drifting is present in 
our data for PSR~B0751+32.

\section{Discussion} \label{sec:disc}

We analyzed the nulling behaviour of PSR~B0751+32 using high-sensitivity FAST observations 
at 1250~MHz. We confirm periodic nulling, with a nulling fraction of \(35.1\% \pm 0.6\%\) from the 
mixture-model analysis, and find that the modulation varies over time. The mean profile shows an 
asymmetric two-component structure, with a brighter, narrower leading component. Pulse-energy 
analysis indicates that both components are stable at the start of the burst state but gradually 
weaken toward its end, with the trailing component more affected. We also find no evidence of 
subpulse drifting, in contrast to earlier claims. 

The nulling periodicity of PSR~B0751+32 clearly demonstrates temporal variability, as the
 periods derived from the two observing sessions differ significantly (Table~\ref{tab:period}).
These observations are consistent with a previous report on PSR~J1136+1551, in which the 
nulling periodicity was found to vary between different observing sessions \citep{hr07}. 
The dynamic power spectrum shown in Figure~\ref{fig:state_spec} 
reveals diffusely distributed modulation features, providing visual evidence for the temporal 
variability of the nulling periodicity within a single observation. Such intra-observation 
variability suggests that the modulation mechanism is sensitive to short-timescale changes in the 
emission conditions. This variability challenges earlier models that attributed 
periodic nulling to a fixed 
geometric effect, such as an empty line of sight traversing gaps between emitting sub-beams 
\citep{rw08}. Instead, it supports the hypothesis that nulling periodicity is influenced by 
changes in the pulsar's magnetosphere or the emission mechanism, which may be subject to 
environmental or intrinsic changes. 

The partially filled rotating beam model remains a plausible mechanism, particularly if the 
central feature of PSR~B0751+32 is attributed to inner-cone rather than core emission. However, an 
inner-cone geometry typically implies that drifting should be observed in both the inner and outer 
cones, showing the same $P_{3}$ values (e.g., \citealt{hw80}). 
The absence of any drifting features in our data therefore argues against this assumption, 
or at least suggests that a purely geometric interpretation is unlikely to be sufficient.

The evolution of pulse energy during state transitions may provide valuable insights into the 
underlying emission process. A gradual decline in pulse energy prior to the null state in 
PSR~B0751+32 --- a phenomenon also reported in other pulsars \citep{tlw+25, rwyw22, wwy+16}, 
particularly in the trailing component --- indicates that the cessation of emission is likely 
a decay process rather than an abrupt switch-off. Such behaviour may plausibly arise from a 
gradual depletion of the magnetospheric plasma supply or from a slow change in the coherence 
conditions necessary for radio emission. The slower fall-off of the leading component further 
indicates that the leading and trailing components may arise from different regions within 
the magnetosphere or be driven by distinct physical mechanisms. 

Based on observations at 430~MHz, \citet{bac81} reported that PSR~B0751+32 shows a tendency for subpulse 
drifting. Subsequent analyses by \citet{wes06,wse07} revealed spectral features with periodicities 
of $60 \pm 20\ P$ in the leading component and $70 \pm 10\ P$ in the trailing component. 
The detection in the leading component in particular is consistent 
with the presence of subpulse drifting.
The absence of subpulse drifting in our high-sensitivity FAST data is noteworthy. 
This discrepancy may be attributed to several factors. First, subpulse drifting is known to 
be frequency-dependent \citep{wbs81,smk05}, and the transition from conal to core-dominated emission at higher 
frequencies could suppress drifting behaviour. However, the fact that \citet{wes06} employed an 
observing frequency of 1.4~GHz, which is very close to ours, renders this explanation less probable.
Second, the pulsar may undergo mode changes, 
where drifting is present only in certain emission states. Our observations, conducted at 1250 MHz, 
might have captured the pulsar in a non-drifting mode. Alternatively, the high sensitivity of FAST may 
have revealed that the apparent drifting reported earlier was actually a manifestation of pronounced periodic 
nulling --- a phenomenon that can mimic drifting features in observations of lower sensitivity. 
This explanation is the most plausible. 

A key distinction between subpulse drifting and periodic amplitude modulation or periodic nulling is that the 
drifting periodicities ($P_3$) have been found to be inversely correlated with $\dot{E}$~\citep{bmm+16,bmm+19}, 
whereas the periodicities in the latter phenomena have shown no significant correlation with $\dot{E}$. If the 
subpulse drifting reported by \citet{bac81} and \citet{wes06,wse07} is genuine, the relatively large $P_3$ of 
PSR~B0751+32 would place it well outside the locus of typical subpulse-drifting pulsars in the $P_3$--$\dot{E}$ 
diagram, thereby challenging the established inverse $P_3$--$\dot{E}$ relation. However, our results indicate 
that PSR~B0751+32 exhibits periodic nulling without subpulse drifting, and thus would not affect the inverse 
$P_3$--$\dot{E}$ correlation observed among subpulse-drifting pulsars.

\section{Summary} \label{sec:summary}

In comparison with earlier studies \citep{bac81,wes06,wse07,hr09}, our new findings in this paper are 
summarised as follows:

1. Using high-sensitivity observations together with a new estimation method, we obtain a more accurate NF of $35.1\% \pm 0.6\%$.

2. We identify clear intra-observation evolution of periodic nulling (Figure~\ref{fig:state_spec}), 
which challenges fixed-geometry interpretations and instead supports a time-variable plasma supply within 
the magnetosphere.

3. Pulse-energy decay behaviour: both components systematically weaken toward the end of each burst 
(Figure~\ref{fig:var_around_null}), rather than disappearing abruptly.

4. Revision of previous claims: we detect no evidence for subpulse drifting (Figure~\ref{fig:2dfs}), 
implying that previously reported “drifting features” were in fact artifacts introduced by periodic nulls under low-sensitivity observations.

In conclusion, our study highlights the complex and dynamic nature of periodic nulling in PSR~B0751+32. 
The temporal evolution of the nulling periodicity and the absence of subpulse drifting underscore the 
need for multi-epoch, multi-frequency observations to fully understand the magnetospheric processes 
responsible for these phenomena. The high sensitivity of FAST has enabled a detailed characterization of the 
nulling behaviour, which may aid future investigations of similar pulsars.

\section*{Acknowledgements}

This work is supported by 
the National Key R\&D Program of China (No. 2022YFC2205201), 
the National Natural Science Foundation of China (NSFC) 
project (No. 12273100, 12288102), the Tianshan Talent 
Training Program (No. 2024TSYCCX0073, 2023TSYCTD0013), 
the CAS project (No. JZHKYPT-2021-06), 
the Major Science and Technology Program of Xinjiang Uygur 
Autonomous Region (No. 2022A03013-4), the Natural Science Foundation 
of Xinjiang Uygur Autonomous Region (No. 2022D01D85). 
This research is partly supported by
the Operation, Maintenance and Upgrading Fund for Astronomical 
Telescopes and Facility Instruments, budgeted from
the Ministry of Finance of China (MOF) and administrated
by the CAS. This work made use of the data from FAST 
(Five-hundred-meter Aperture Spherical radio Telescope). FAST is 
a Chinese national mega-science facility, operated by National 
Astronomical Observatories, Chinese Academy of Sciences.

\section*{Data Availability}

The data underlying this article will be shared on reasonable request to the 
corresponding author.











\bsp	
\label{lastpage}
\end{document}